\documentclass[sigconf]{acmart}
\usepackage{xcolor}
\usepackage{caption}
\usepackage{comment}
\definecolor{uiblue}{HTML}{3B89F0}
\definecolor{uigreen}{HTML}{5EB761}

\AtBeginDocument{%
  }

\setcopyright{acmlicensed}
\copyrightyear{2026}
\acmYear{2026}
\acmDOI{10.1145/3772363.3798685}
\acmConference[CHI'26]{April 11--18, 2026}{Barcelona, Spain}

\acmISBN{978-1-4503-XXXX-X/2018/06}

\begin{document}

\title{"Who wants to be nagged by AI?": Investigating the Effects of Agreeableness on Older Adults’ Perception of LLM-Based Voice Assistants’ Explanations}

\author{Niharika Mathur}
\email{nmathur35@gatech.edu}
\orcid{0000-0002-3969-7787}
\affiliation{%
  \institution{Georgia Institute of Technology}
  \city{Atlanta}
  \state{Georgia}
  \country{USA}
}

\author{Hasibur Rahman}
\email{rahman.has@northeastern.edu}
\orcid{0009-0008-0938-1504}
\affiliation{%
  \institution{Northeastern University}
  \city{Boston}
  \state{Massachusetts}
  \country{USA}
}

\author{Smit Desai}
\authornote{Corresponding author}
\email{sm.desai@northeastern.edu}
\orcid{0000-0001-6983-8838}
\affiliation{%
  \institution{Northeastern University}
  \city{Boston}
  \state{Massachusetts}
  \country{USA}
}

\renewcommand{\shortauthors}{Mathur et al.}
\renewcommand{\shorttitle}{Agreeableness in LLM-VAs}

\begin{abstract}
LLM-based voice assistants (VAs) increasingly support older adults aging in place, yet how an assistant's agreeableness shapes explanation perception remains underexplored. We conducted a study (N=70) examining how VA agreeableness influences older adults' perceptions of explanations across routine and emergency home scenarios. High-agreeableness assistants were perceived as more trustworthy, empathetic, and likable, but these benefits diminished in emergencies where clarity outweighed warmth. Agreeableness did not affect perceived intelligence, suggesting social tone and competence are separable dimensions. Real-time environmental explanations outperformed history-based ones, and agreeable older adults penalized low-agreeableness assistants more strongly. These findings show the need to move beyond a \textit{one-size-fits-all} approach to AI explainability, while balancing personality, context, and audience.
\end{abstract}




\maketitle

\section{Introduction}

Voice Assistants (VAs) are increasingly embedded in home environments, where they support users by providing timely reminders \cite{lee2015sensor, mathur2022collaborative} and safety alerts \cite{ejupi2016kinect}. In such environments, the increasing conversational fluency of LLM-based VAs or LLM-VAs introduces parallel challenges for how users understand and interpret them. For older adults aging in place, explanations play a critical role in helping them make sense of \textit{why} an AI acted in a certain way (e.g., why a reminder was issued at a specific moment) \cite{gleaton2023understanding, mathur2023did}. Despite recognizing the importance of AI explanations in calibrating trust, relatively little is known about how explanations are \textit{experienced} by users interacting with conversational, LLM-VAs in real-world settings \cite{ehsan2020human, miller2019explanation, ehsan2022human}. Research in explainable AI (XAI) has largely focused on mechanistic interpretations of AI models, often emphasizing internal model logic and decision factors \cite{miller2019explanation}. However, in VAs, explanations are delivered as conversational responses through natural language, and hence, LLM personality traits can affect how users interpret and rely on them. As a result, it is just as important to examine \textit{how} an explanation is communicated conversationally in real-world settings, than it is to focus on technical interpretations of explanations \cite{mathur2025sometimes, he2025conversational}.     

At the same time, research in Conversational User Interfaces (CUIs) has shown that most LLM-VAs are often designed to exhibit high agreeableness to promote user engagement \cite{rahman2025vibe, malmqvist2025sycophancy}. Psychometric research defines \textit{Agreeableness} as a personality trait of being kind, sympathetic, cooperative, warm, honest, and considerate \cite{john1999big}. However, this high agreeableness conversational behavior can also unintentionally reduce critical scrutiny and promote overreliance \cite{sun2025friendly}, with users relying on the LLM-VA because it sounds polite \cite{pak2025polite}. In such situations, when an LLM-VA delivers explanations with a specific level of agreeableness, their persuasive impact may be amplified -- highlighting the need to examine how conversational and personality traits in LLM-VAs shape perceptions of their explanations. To examine this, we conducted a controlled experimental study (N=70) examining how variation in a fictional LLM-VA’s agreeableness influences older adults’ perceptions of explanations across everyday in-home scenarios. We compare explanations grounded in
prior conversational history and real-time environmental data across two task contexts, and consider how older adults' own agreeableness shapes their responses to the VA. Our findings contribute evidence that high agreeableness in LLM-VAs improves older adults’ trust, empathy and likeability towards AI-generated explanations without affecting perceived intelligence. We further show that situational task context also matters, highlighting the need to design LLM-VAs that prioritize personality-aware explanation design in aging in place settings.

\section{Related Work and Background}


\textbf{\textit{LLM-VAs and Explanations.}} VAs have traditionally been used in assistive settings to provide routine reminders \cite{lee2015sensor, mathur2022collaborative}, safety alerts \cite{ejupi2016kinect, mohan2024artificial, wu2018understanding}, and social and conversational support \cite{el2020multimodal} to users. More recently, the new generation of LLM-based VAs now allow for more flexible and context-aware dialog, capable of referencing prior conversational history, and embodying distinct personality cues. However, while transformative, LLMs also introduce parallel challenges concerning their \textbf{explainability} \cite{enam2025artificial, gleaton2023understanding, mathur2023did}. Emerging research on LLMs has identified several limitations related to opacity \cite{manche2022explaining}, hallucinations \cite{orgad2024llms, perkovic2024hallucinations}, lack of transparency in training data \cite{ehsan2024human}, and LLM personality traits such as sycophancy \cite{carro2024flattering, sun2025friendly}, all of which undermine the reliability and trustworthiness of LLM-generated explanations. 

When people interact with VAs, they often seek social engagement \cite{namvarpour2025art, pradhan2018accessibility}, and evaluate VAs based on its personality cues \cite{desai_examining_2024, gilad2021effects, zhou2019trusting}. Among the many personality traits that shape VA perceptions, research has shown that \textit{agreeableness} is particularly consequential. Agreeableness, one of the Big Five personality traits\footnote{
The Big Five personality framework (also known as OCEAN) is a human personality classification model in psychology, comprising five broad dimensions: openness, conscientiousness, extraversion, agreeableness, and neuroticism (OCEAN) \cite{john1991big, john1999big, goldberg1993structure}.}, has shown to significantly influence user perceptions of a VA's likeability \cite{volkel2021examining, chin2024like}, intention to adopt \cite{stein2024attitudes, park2022likes}, and trust \cite{rahman2025vibe}, particularly in contexts where VAs provide assistive support like aging in place \cite{chin2024like}. While many studies have examined this influence of agreeableness on user perception, the increased conversational depth of LLMs necessitates this inquiry to explanations as well, given their crucial role in calibrating user trust \cite{mathur2025sometimes, kim2024m}, motivating our first research question: \textit{RQ1: How does variation in an LLM-VA's agreeableness influence older adults’ perception of its explanations?}


\textbf{\textit{Human-Centered AI Explanations for Aging in Place}}. Prior work in HCI has highlighted the need to design explanations rooted in context, which involves leveraging resources beyond the AI model itself (such as prior interactions, domain knowledge, and contextual information) to generate explanations aligned with user goals and expectations \cite{liao2021question, ehsan2024xai}. These considerations are particularly salient for aging in place, where interactions with VAs are shaped by  cognitive, social and environmental factors \cite{bagnall2006older}. When older adults interact with AI systems, they frequently draw on familiar elements of their sociotechnical environments and their interaction history to interpret AI's behavior, such as referencing physical objects, or prior conversations with the AI \cite{jacelon2013older, pradhan2021entanglement, mathur2025feels}, and explanations grounded in such contextual information significantly improve understanding \cite{kofod2007explanations, das2023explainable, baruah2024brief}. Building on this, \citet{mathur2024categorizing, mathur2025sometimes} identify \textit{conversational history} and \textit{real-time environmental data} as two key information sources that can be leveraged in smart home environments to generate explanations, a categorization that we adopt in our study to understand older adults' perception of LLM-VA explanations. To further examine how this perception varies across tasks with differences in urgency and risk, as these are significant predictors of human-AI interaction \cite{salimzadeh2024dealing, xu2025enhancing}, we ask our second research question: \textit{RQ2: How is older adults’ perception of an LLM-VA’s explanations influenced by situational context (routine vs. emergency) and explanation type?} Finally, recognizing that older adults are not a homogeneous user group, and that their own personality traits, like agreeableness, also shape their perceptions of technology \cite{chin2024like, volkel2021examining}, we ask our final research question: \textit{RQ3: How does older adults’ agreeableness influence their perception of the LLM-VA’s explanations?} 

\section{Method}


\textbf{\textit{Study Overview.}} Participants interacted with \textit{Robin}, a fictional LLM-VA designed to support older adults, using vignette-based interactive storyboards. We employed a 2 x 2 x 2 mixed factorial design, manipulating explanation type (conversational history vs. real-time environmental data), LLM-VA personality (agreeableness: high vs. low), and interaction context (routine vs. emergency). Agreeableness was manipulated as a \textit{between-subjects} factor, while explanation type and interaction context were manipulated \textit{within subjects}, enabling examination of LLM-VA's personality effects (RQ1), contextual and explanation-type effects (RQ2), and individual differences among older adults (RQ3).

To differentiate situational contexts in our study, we selected two assistive contexts: 
 \textbf{routine reminders} (low-risk, non-urgent support for daily activities) and \textbf{emergency alerts} (high-risk, time-sensitive notifications such as safety hazards), informed by prior work on smart home assistance and conversational AI use for aging \cite{mynatt2001digital, das2023explainable, mathur2022collaborative, pradhan2020use}. Within each context, participants interacted with a set of scenarios, presented using visual storyboards depicting a box-like VA providing a reminder or alert, followed by an explanation. The explanation types were grounded in the categorization by Mathur et al. \cite{mathur2024categorizing, mathur2025sometimes}. We operationalized two explanation types: \textbf{conversational user history-based explanations (UH)}, which referenced a user’s prior interactions with \textit{Robin}, and \textbf{real-time environmental explanations (ENV)}, which drew on sensor-based data reflecting immediate, real-time situational factors \cite{mathur2024categorizing, mathur2025sometimes}. Explanations were generated using GPT-5.0 based on prompts specifying the scenario, context, and information source, and were conditioned to match the agent's agreeableness level (high vs. low) using the Trait Modulation Key (TMK) prompting framework \cite{rahman_vibe_2025}, and then manually reviewed to ensure consistency and clarity.

\textbf{\textit{User Interface.}} Storyboard-based studies can force users to infer unspecified interaction from a narrative artifact, limiting how well dynamic interaction is conceptualized and communicated \cite{kim_sketchstudio_2018, benford_interaction_2009}, so \citet{dow_external_2006} argue for additional tools to support storyboards. Hence, we embedded audio clips into an interactive storyboard rather than using audio-only stimuli as in prior VA studies \cite{desai_examining_2024, pias_impact_2024, jamshed_designing_2025}, making it easier for participants to imagine how the exchange would unfold in context. We implemented this approach in a web-based, interactive storybook UI (Fig. 1) with two side-by-side panels. The left panel provided context, and the right panel displayed the follow-up explanation. Interaction followed a fixed A$\rightarrow$B$\rightarrow$C flow. In (A), participants clicked the \textcolor{uiblue}{\Large$\blacktriangleright$} button on \textit{Robin} to hear the reminder or alert, after which a \textcolor{uigreen}{\Large$\blacksquare$} speech bubble displayed the transcript. In (B), a second \textcolor{uiblue}{\Large$\blacktriangleright$} button appeared on the right panel to play the explanation, followed by its \textcolor{uigreen}{\Large$\blacksquare$} transcript bubble. In (C), the \textcolor{uigreen}{\Large$\blacksquare$} transcript was shown after the explanation was played. The storyboards used in the study were created using Gemini Language Model.

\begin{figure*}[ht]
    \centering
    \includegraphics[width=\linewidth]{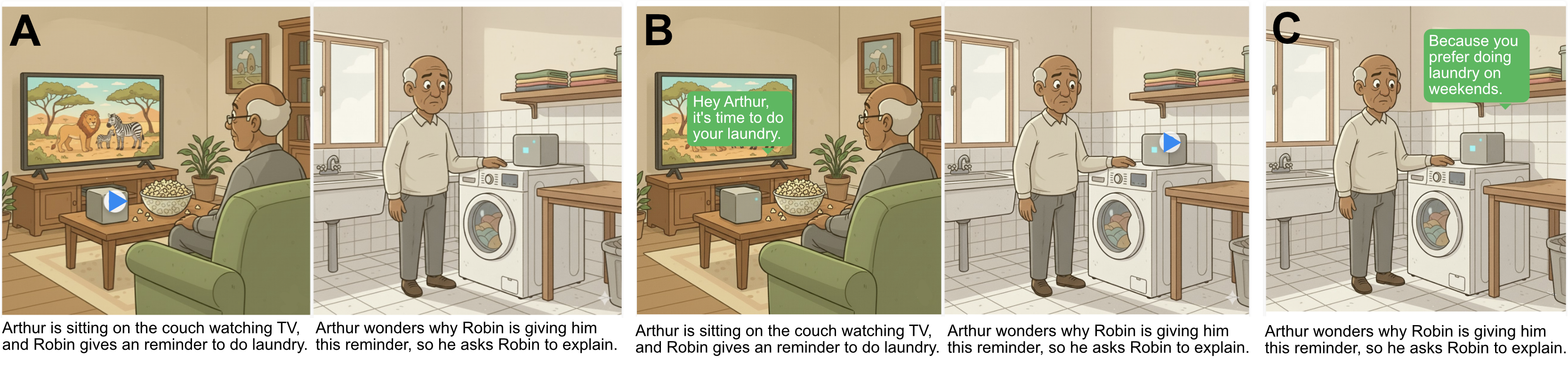}
    \caption{Interactive storyboard UI and sequence (A$\rightarrow$B$\rightarrow$C). Participants clicked the \textcolor{uiblue}{\Large$\blacktriangleright$} button to play the reminder or alert (A) and then the explanation (B). After each audio clip, a \textcolor{uigreen}{\Large$\blacksquare$} speech bubble displayed the transcript (B)(C).}
    \label{fig:xai}
\end{figure*}

\textbf{\textit{Measurements. }}To evaluate users' perceptions of \textit{Robin}, we employed seven validated measures. 6-item \textbf{Empathy} was rated on a 101-point slider using the Perceived Empathy of Technology Scale (PETS)  \cite{schmidmaier_perceived_2024}, and 2-item \textbf{intention to adopt} on a 7-point Likert scale by \citet{moussawi_how_2021}.
Perceived \textbf{likeability} and \textbf{intelligence} were assessed using the Godspeed questionnaire (5 items each) \cite{bartneck_measurement_2009}, measured with 5-point semantic differential scales. The remaining measures used 5-point Likert scales: \textbf{trustworthiness} from the Trust in Automation (TiA) questionnaire (5 items) \cite{jung_great_2022, jian2000foundations}, explanation \textbf{satisfaction} from  \citet{hoffman_measures_2023} (7 items), and \textbf{reliance} (5 items). To measure participants' agreeableness, we used 4 domain-specific items from the Mini-IPIP\footnote{The Mini-IPIP (Mini-International Personality Item Pool) is a questionnaire used to measure the Big Five personality traits in a reliable way.} \cite{donnellan_mini-ipip_2006}, rated on a 5-point Likert scale. We also included one manipulation-check item to validate the perceived agreeableness of \textit{Robin} using a 5-point Likert scale. The full batch of storyboards, questionnaires, the prompts used, data and results can be found in the supplementary files. 

 \subsection{Participants}
We recruited 70  U.S.-based older adults through Prolific\footnote{https://www.prolific.com/}, ranging from 60 to 83 years (\(M = 65.97\), \(SD = 5.28\)),  with 35 participants assigned to each experimental condition. 53 participants identified as female, 16 as male, and 1 as other. With respect to prior experience with both VAs and LLM-VAs, 62 participants reported having more than a little experience using them. Participant compensation exceeded the  U.S. federal minimum wage guidelines. There was no significant difference in participants' agreeableness across the experimental condition (\(M = 4.25\), \(SD = 0.67\)).

\subsection{Procedure}

After providing consent, participants completed a demographic survey and the Mini-IPIP personality inventory to assess individual differences in agreeableness. They were then introduced to \textit{Robin} and the interactive storyboards, along with a description of \textit{Robin}’s capabilities.
Following the \textit{between-subjects} manipulation of \textit{Robin}'s agreeableness, participants were assigned to either a high or low-agreeableness version of \textit{Robin} (counterbalanced). Within this condition, they viewed storyboard-based scenarios across two assistive contexts (routine reminders and emergency alerts). For each context, participants saw four scenarios per explanation type (counterbalanced between conversational history (UH) and real-time environmental (ENV) explanations), and rated the assistant after each block, including a single-item manipulation-check of perceived agreeableness, attention checks, and an open-ended text field for qualitative reflections.

\section{Results}
To assess the success of our agreeableness manipulation using the TMK prompting framework \cite{rahman_vibe_2025}, we compared responses to the manipulation-check question on assistant agreeableness between the low-agreeableness (LA) and high-agreeableness (HA) versions of \textit{Robin}. Participants rated HA \textit{Robin} as significantly more agreeable than LA \textit{Robin} (HA: $M = 4.171$ vs.\ LA: $M = 3.229$), $t = -4.093$, $p = .0001$. All seven measures demonstrated good to excellent internal consistency, with $\alpha = .883$--$.985$ and $\omega = .896$--$.985$. Additionally, we conducted an exploratory factor analysis of the reliance scale, which supported a one-factor solution (eigenvalue $= 3.554$), explaining $64.5\%$ of the variance, with all items loading strongly ($>.64$).
Because none of the outcomes were normally distributed, we used nonparametric tests.
 
\begin{figure*}[ht]
    \centering
    \includegraphics[width=1\linewidth]{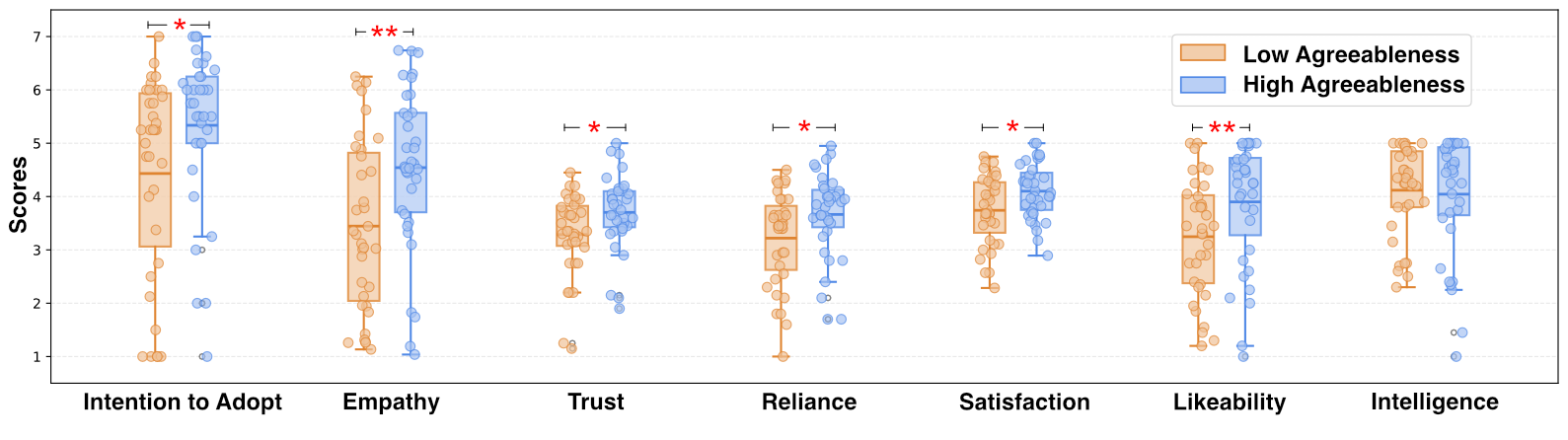}
    \caption{Ratings for LA versus HA \textit{Robin} across seven outcomes. Asterisks mark significant LA–HA differences from Kruskal-Wallis tests (\textcolor{red}{*} $p<.05$, \textcolor{red}{**} $p<.01$). Intention to adopt and empathy are shown on 1 to 7 scales, with empathy scores rescaled for this visualization only. Trust, reliance, satisfaction, likeability, and intelligence are shown on 1 to 5 scales.}

    \label{fig:result}
\end{figure*}

\subsection{LLM-VA Agreeableness and Explanation Perception (RQ1)}
To examine whether the LLM-VA’s agreeableness shaped older adults’ perceptions of its explanations (RQ1), we compared ratings between LA \textit{Robin} and HA \textit{Robin} using Kruskal-Wallis tests, which are equivalent to Mann-Whitney $U$ tests for two groups.  We report effect sizes as the absolute rank-biserial correlation, $|r|$. Participants reported higher \textit{intention to adopt} for HA (Mdn = 5.75) than LA (Mdn = 5.25), $H = 5.299$, $p = .021$, $|r| = .275$ (small). HA \textit{Robin} was also perceived as more \textit{empathetic} (HA Mdn = 59.32 vs.\ LA Mdn = 38.16), $H = 7.394$, $p = .006$, $|r| = .325$ (medium), and more \textit{likeable} (HA Mdn = 4.25 vs.\ LA Mdn = 3.40), $H = 7.188$, $p = .007$, $|r| = .320$ (medium). We also observed significant, small-sized differences favoring HA \textit{Robin} for \textit{trust} (HA Mdn = 3.75 vs.\ LA Mdn = 3.35), $H = 5.666$, $p = .017$, $|r| = .285$, \textit{reliance} (HA Mdn = 3.85 vs.\ LA Mdn = 3.45), $H = 5.013$, $p = .025$, $|r| = .268$, and \textit{satisfaction} (HA Mdn = 4.07 vs.\ LA Mdn = 3.75), $H = 4.624$, $p = .032$, $|r| = .257$. In contrast, effects on perceived \textit{intelligence} was not significant, $H = 0.134$, $p = .715$, $|r| = .044$. \textbf{To answer RQ1, as shown in Fig.~\ref{fig:result}, HA \textit{Robin} led older adults to rate its explanations more favorably across most outcomes (e.g., intention to adopt, empathy, likeability, trust, reliance, satisfaction), with no difference in perceived intelligence.}

\subsection{Context and Explanation Type Effects (RQ2)}
We explored whether \textit{explanation context} (reminders, Ctx1, vs.\ alerts, Ctx2) and \textit{explanation type} (UH vs.\ ENV) moderated the effect of \textit{Robin}’s agreeableness (LA vs.\ HA) on explanation perceptions. For the between-subject effect, Kruskal-Wallis tests showed that HA outperformed LA across more measures in Ctx1 (\textit{intention to adopt, empathy, trust, reliance, satisfaction, likeability}; all $p < .05$) than in Ctx2 (only i\textit{ntention to adopt, empathy, likeability}; all $p < .05$). Similarly, HA's advantage was broader for UH explanations (all measures except \textit{intelligence},  $p < .05$) than for ENV explanations (\textit{intention to adopt, empathy, trust, likeability};  $p < .05$). For the within-subject effect, Wilcoxon tests indicated that Ctx1 produced higher \textit{intention to adopt} ($W=337.0$, $p=.0253$) and \textit{trust} ($W=729.5$, $p=.0377$)) than Ctx2. Similarly, ENV explanations outperformed UH on \textit{reliance, satisfaction, likeability, and intelligence} (all $p < .05$). All other within-subject comparisons were not significant. \textbf{In addressing RQ2, routine contexts showed higher \textit{intention to adopt and trust} compared to the emergency context. ENV explanations were generally preferred over UH, resulting higher reliance, satisfaction, likeability, and intelligence. HA \textit{Robin}'s advantage over LA was strongest in routine contexts and UH explanations, weakened in emergency contexts.}


\subsection{User Agreeableness and Explanation Perception (RQ3)}

For LA \textit{Robin}, participant agreeableness showed significant negative correlations with \textit{intention to adopt}, $\rho=-.42$, $p=.012$, \textit{trust}, $\rho=-.36$, $p=.034$, and \textit{likeability}, $\rho=-.50$, $p=.002$. Correlations with \textit{empathy} ($\rho=-.32$, $p=.062$), \textit{reliance} ($\rho=-.32$, $p=.063$), \textit{satisfaction} ($\rho=-.28$, $p=.100$), and \textit{intelligence} ($\rho=-.30$, $p=.085$) were not significant. For HA \textit{Robin}, participant agreeableness was not significantly correlated with any outcomes. \textbf{In response to RQ3, we found an interesting association between older adults’ agreeableness and their perceptions of \textit{Robin}’s explanations: more agreeable older adults penalized the LA \textit{Robin} more.}

\subsection{Qualitative perceptions}
 
Open-ended responses helped contextualize the quantitative differences between HA and LA \textit{Robin} by revealing how older adults interpreted explanations at a social and emotional level. Participants interacting with the LA \textit{Robin} often described its explanations as abrupt, unsupportive, or judgmental, with wording perceived as discouraging or containing ``\textit{unnecessary digs},'' prompting comments such as ``\textit{who wants to be nagged by AI?}''  In contrast, explanations from the HA \textit{Robin} were commonly described as friendly, thoughtful, and polite, aligning with higher ratings for empathy and satisfaction. Across conditions, participants attended less to the factual content of explanations than to whether \textit{Robin}'s tone met expectations of emotional support, particularly in routine reminder contexts. 

\section{Discussion}

\textbf{\textit{Design Considerations.}} Our findings extend prior work on conversational AI personality and explainability, suggesting five design considerations (DCs). \textbf{DC1:} Consistent with \citet{volkel2021examining}, who found users prefer agreeableness contextually rather than universally, our results demonstrate that effective explanations require sensitivity to context, grounding, and audience, not a fixed conversational style. \textbf{DC2:} High agreeableness improved trust, empathy, and adoption in routine contexts, but these benefits diminished in emergencies, supporting \citet{polite}'s findings that agreeableness effects are task-dependent and suggesting clarity may outweigh warmth in high-stakes scenarios. \textbf{DC3:} Agreeableness increased likeability without affecting perceived intelligence, indicating these are separable design dimensions. This extends \citet{khadpe2020conceptual}'s work on conceptual metaphors and suggests designers can modulate warmth independently of competence signaling, critical for avoiding sycophancy concerns in LLMs \citep{sun2025friendly}. \textbf{DC4:} Real-time environmental explanations were perceived as more credible than history-based ones, echoing \citet{mathur2025sometimes}'s emphasis on situationally anchored information for older adults and suggesting data-driven grounding may better support trust calibration than personalization alone. \textbf{DC5:} More agreeable older adults penalized low-agreeableness assistants more strongly, extending personality-matching research \citep{volkel2021examining, rahman_vibe_2025, chin2024like} to the explanation domain. Collectively, these findings position AI explanation design as a joint problem of content, delivery, and audience, reinforcing that \textit{one-size-fits-all} approaches are insufficient for older adults across diverse assistive scenarios.

\textbf{\textit{Limitations and Future Work.}} In this study, we examined a single personality dimension, agreeableness, using a binary manipulation that enabled controlled comparison but did not capture the full range of personality expression in conversational AI. Future work should examine additional traits, such as emotional stability and conscientiousness. Our storyboard-based design relied on a single synthetic voice, limiting insight into how vocal tone and prosody interact with personality, as well as longitudinal dynamics or the impact of repeated system failures in-the-wild. Finally, our scenarios were presented in a controlled setting, and in-situ deployments are needed to understand how these effects evolve with sustained real-world use.

\section{Conclusion}

This study examined how an LLM-based VA’s agreeableness shaped older adults’ perceptions of AI explanations. High agreeableness increased trust, empathy, and adoption without raising perceived intelligence, while low agreeableness was penalized, especially by agreeable users. Explanation perceptions were context-sensitive and depended more on grounding than personality alone, indicating that effectiveness emerged from the interaction of assistant agreeableness, context, and user traits.

\begin{acks}
  We would like to thank the reviewers and the members of Conversational Human-AI Interactions (CHAI) Lab for their feedback and thoughtful engagement with this work. 
\end{acks}

\bibliographystyle{ACM-Reference-Format}
\bibliography{cites}

\appendix

\end{document}